\documentclass[aps,pra,a4paper,showpacs,twocolumn]{revtex4}

\usepackage[latin1]{inputenc}
\usepackage{amsmath}
\usepackage{amsfonts}
\usepackage{amssymb}
\usepackage{bm}
\usepackage{graphicx}
\usepackage{epsfig}
\usepackage{subfigure}
\usepackage{array}
\usepackage{subeqnarray}

\begin{document} 
\renewcommand{\>}{\rangle}
\newcommand{\<}{\langle }
\title{Arbitrary state controlled-unitary gate by adiabatic passage}
\date{\today}
\pacs{03.67.Lx, 32.80.Qk}
\author{X. Lacour$^{1}$, N. Sangouard$^{2}$, S. Gu\'{e}rin$^{1}$, and H.R. Jauslin$^{1}$}
\affiliation{$^{1}$Laboratoire de Physique, Universit\'e de Bourgogne,
UMR CNRS 5027, BP 47870, 21078 Dijon Cedex, France \\
$^{2}$ Fachbereich Physik, Universit\"at Kaiserslautern,\\
67653, Kaiserslautern, Germany}
\begin{abstract}
We propose a robust scheme involving atoms fixed in an optical cavity
to directly implement the universal controlled-unitary gate.
The present technique based on adiabatic passage uses novel dark states well suited for the controlled-rotation operation.
We show that these dark states allow the robust implementation of a gate that is a  generalisation of the controlled-unitary gate to the case where the control qubit can be selected to be an arbitrary state.
This gate has potential applications to the rapid implementation of quantum algorithms such as of the projective measurement algorithm.
This process is decoherence-free since excited atomic states and cavity modes are not populated during the dynamics.

\end{abstract}
\maketitle
\section{Introduction}
The realization of a universal quantum circuit is a great
challenge in quantum information science.
It is known that an arbitrary quantum computation can be performed by combining quantum gates, i.e.
unitary operators acting on qubits,
that form a \emph{universal set}.
We distinguish two types of universal sets. The first one is composed of a general one-qubit gate [corresponding to a general operator $U$ of
$\hbox{SU}(2)$] and a two-qubit entangling gate~\cite{Bremner}\,; the second type is composed by a single kind of gates, called \emph{universal gates}, like the controlled-unitary gate (C\hbox{-}U)~\cite{Chuang}.

In order to make quantum computations, the implementations of
these quantum gates have to be robust, i.e. they have to be insensitive to
fluctuations or to partial knowledge of experimental parameters.
Furthermore they have to be insensitive to decoherence effects,
such as spontaneous emission. Those conditions can be fulfilled if the qubit is encoded in
atomic metastable states, and if the gates
are implemented by adiabatic passage along dark states, i.e.
instantaneous eigenstates with time independent eigenvalue (equal to the energy of the ground states) and with zero projection on the excited states.

However, adiabaticity is not sufficient to insure the robustness of certain quantum gates. 
The parameters that determine the action of the gates on qubits, like the argument of the rotation gate or the phase of the controlled-phase gate,
have to be controlled with high accuracy to perform computation~\cite{Preskill}.
We therefore have to avoid the use of the non-robust dynamical phases, depending on the area under the adiabatic pulses, and of geometric phases~\cite{Duan} that require the control of a loop in the parameter space.
An alternative technique consists in using elliptic polarisation and static phase difference of lasers, which can be easily controlled experimentally.
Following this idea, the implementation of a general single qubit gate based on fractional stimulated Raman adiabatic passage (f-STIRAP)~\cite{bergmann2,vitanov} in a tripod-type system~\cite{unanyan} has been proposed in Ref.~\cite{kis}. A multi-controlled-unitary gate acting on qubits fixed in an optical cavity has been proposed in Ref.~\cite{Goto}, but with an undesirable phase gate that has to be compensated. The latter proposition is based on the two-qubit adiabatic transfert described in Ref.~\cite{Pellizari} and f-STIRAP.

Since in experimental implementations of quantum computations the errors grow with the number of quantum gates involved, it is advantageous to implement directly certain gates instead of relegating them to a combinaison of elementary gates, as illustrated in Ref.~\cite{Sangouard} with the direct implementation of the SWAP gate. There is a double advantage\,: to reduce the errors with a smaller number of gates and to decrease decoherence effects by reducing the computational time.

In this paper we propose a direct implementation by adiabatic passage along dark states of a \textit{arbitrary state controlled-unitary gate} using only seven pulses. This gate can be writen as
\begin{equation}
\left[
\begin{matrix}
\openone & 0 \\
0 & U\\
\end{matrix}
\right]
\end{equation}
in the basis $\lbrace|\phi_{nc}0\>, |\phi_{nc}1\>,|\phi_{c}0\>,|\phi_{c}1\>\rbrace$, where $U$ ( respectively $\openone$) are an unitary (the identity) operator of $\hbox{SU}(2)$, $\phi_{c}$ an arbitrary control state of the first qubit, $\phi_{nc}$ its orthogonal state.
 This process is based on dark states generalising those of f-STIRAP and of Ref.~\cite{Pellizari}, and particularly adapted for a two-qubit controlled-rotation operation. This gate has potential applications for the rapid realisation of quantum algorithms. We show for instance that it allows a direct implementation of the projective measurement algorithm.
The paper is organised as follows. The system is introduced in section~\ref{System}. The definition and the dynamics of the gate are shown in section~\ref{Dynamics}. Section~\ref{Numeric} is devoted to the numerical demonstration, and section~\ref{End} presents some conclusions.

\section{System}\label{System}
\begin{figure}[ht]
\includegraphics[scale=1]{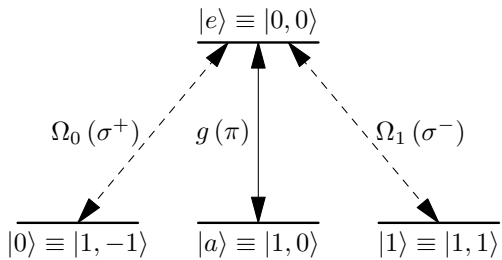}
 \caption{
Schematic representation of a Zeeman sublevel system with the
polarisation of the fields driving each transition associated with the
states $|J,\,M\>$ for $J=0,\,1$. For the generation of
single qubit gates, each transition is driven by laser\,; for the two-qubit states rotation a cavity field drives the transition $|a\>-|e\>$, the others are driven by lasers.}\label{Zeeman}
 \end{figure}

As in Refs.~\cite{Pellizari,Goto,Kapale} we consider a register of qubits fixed in an optical cavity.
Each qubit is encoded in a tripod-type Zeeman system composed of three metastable states and one excited state, and can be addressed individually by laser fields.
The qubits interact with each other through the cavity mode~\cite{Pellizari}.
The single qubit gates can be implemented in this system by coupling the three ground states by lasers~\cite{SingleQG}.
We choose here to identify the Zeeman sublevels $|J=1,\,M=\pm1\>$ to the computational states $|0\>$ and $|1\>$. The ancillary state $|J=1,\,M=0\>$ and the excited state $|J=0,\,M=0\>$ are respectively denoted $|a\>$ and $|e\>$ (see Fig.~\ref{Zeeman}).

During the dynamics, the excited state $|e\>$ is coupled to the computational states
$|0\>$ and $|1\>$ by circularly polarised laser fields of Rabi frequencies
$\Omega_{0}^{(k)}$ and $\Omega_{1}^{(k)}$ (the superscript $k$ labels the atoms),
and to the ancillary state $|a\>$ by the linearly polarised cavity mode of Rabi frequency $g^{(k)}$ which
is time independent. Each field is one-photon resonant, and their
polarisation and frequencies are such that they drive a unique
transition.
The choice of the
polarizations is guided by geometrical constraints, when we
impose that the lasers propagate orthogonally to the cavity axis.
The essential point is that the polarisation of the cavity mode is orthogonal to the plane of the circular polarisation of the lasers.

\section{Dynamics}\label{Dynamics}
\begin{figure}[!htb]
\includegraphics[scale=0.95]{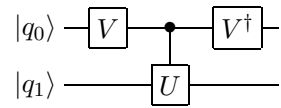}
\caption{Quantum circuit representing the decomposition of the arbitrary state controlled-unitary gate from elementary gates.}
\label{circ1}
\end{figure}

In this section we describe the sequence of pulses that will permit to generate a \emph{arbitrary state controlled-unitary} gate (C$_{\hbox{as}}$-U). 
We recall that a standard controlled-unitary gate yields a unitary operation on the target qubit if the control qubit is in state $|1\>$. 
We define its generalisation as follows. The arbitrary state controlled-unitary gate yields a unitary operation on the target qubit if the control qubit 
is in an arbitrary preselected state that we can choose robustly by the laser pulse parameters. It is equivalent to the quantum circuit represented on Fig.~\ref{circ1}, where $U,V$ are unitary operators and $|q_0\>,|q_1\>$ the control and target qubit.

\subsection{Background}
We use the following notation\,: the states of the system are
written $|s_{1}s_{2}\>|n\>$, where the indices $s_{1},\,s_{2}$
denote respectively the states of the first and second qubit, and
$n$ the photon number state of the cavity-mode. 

In Ref.~\cite{Pellizari}, a robust tool has been established to drive a complete population transfer between two-qubit states $|s_1s_2\>|0\>$.
For instance, if the states $|a\>$ of each atom are coupled by the cavity, then a counter-intuitive pulse sequence $\Omega_{1}^{(2)},\Omega_{0}^{(1)}$  induces the population transfer $|0a\rangle|0\rangle\rightarrow|a1\rangle|0\rangle$.
Such a coherent manipulation of the two-atom state
$|s_{1}s_{2}\rangle|0\rangle$ offers various possibilities for
the implementation of two-qubit quantum gates. This tool is at the heart
of the SWAP gate in Ref.~\cite{Sangouard} and of the CNOT gate in Ref.~\cite{CNOT}.

As STIRAP~\cite{bergmann2} can be extended to
f-STIRAP~\cite{vitanov}, one can extend this process to
the creation of coherent superpositions of the two-atom states.
In this case, we use three laser fields of the form $E_i^{(k)}\cos(\omega t+\phi_i^{(k)})$ [$i=0,1$, $k=1,2$], coupling
respectively the states $|1\rangle$ and $|e\rangle$ of the first
atom, the states $|0\rangle-|e\rangle$ and  $|1\rangle-|e\rangle$
of the second atom.
In the interaction picture and under the rotating wave approximation the Hamiltonian is given by
\begin{eqnarray}\nonumber
H&=&\Omega_{1}^{(1)}e^{-i\phi_1^{(1)}}|e^{(1)}\>\<1^{(1)}|+g^{(1)}\hat{a}|e^{(1)}\>\<a^{(1)}|\\ \nonumber
& + &\Omega_{0}^{(2)}e^{-i\phi_0^{(2)}}|e^{(2)}\>\<0^{(2)}|+
\Omega_{1}^{(2)}e^{-i\phi_1^{(2)}}|e^{(2)}\>\<1^{(2)}|\\ 
&+ &g^{(2)}\hat{a}|e^{(2)}\>\<a^{(2)}|+h.c.
\end{eqnarray}
with $\hat{a}$ the anihilation operator of the cavity mode, $\Omega_i^{(k)}$ the Rabi frequencies associated to the laser amplitudes $E_i^{(k)}$.

The interaction of the second qubit is parametrised by the following laser Rabi frequencies\,:
\begin{subeqnarray}
\Omega^{(2)}_{0}(t)&=&\Omega^{(2)}(t)\,\sin\theta,\\
\Omega^{(2)}_{1}(t)&=&\Omega^{(2)}(t)\,\cos\theta,
\end{subeqnarray}
which can be generated in a robust way using a single laser of appropriate elliptic polarisation. We refer to such a laser as $\Omega^{(2)}$ in what follows.
We define for the second qubit one non-coupled  and three coupled states as
\begin{subeqnarray}
|\Phi_{nc}\>&=&\cos\theta\,e^{i\phi^{(2)}}|1\>-\sin\theta|0\>,\\ 
|\Phi_{c}\>&=&\sin\theta\,e^{i\phi^{(2)}}|1\>+\cos\theta|0\>,\\ 
|\Phi_{c2}\>&=&\sin\theta\,e^{i\phi^{(2)}}|1\>-\cos\theta|0\>,\\ 
|\Phi_{c3}\>&=&\cos\theta\,e^{i\phi^{(2)}}|1\>+\sin\theta|0\>,
\end{subeqnarray}
where $\phi^{(2)}=\phi^{(2)}_{1}-\phi^{(2)}_{0}$.

The hamiltonian admits the following dark states, i.e. instantaneous eigenstates of null eigenvalues and not connected to excited atomic states,
which belong to three orthogonal subspaces\,:
\begin{subeqnarray}
|\Psi_1\rangle  &=&|0\Phi_{nc}\>|0\rangle \,,\\
|\Psi_2\rangle &=&\cos\eta|0\Phi_{c}\rangle|0\rangle-\sin\eta\,e^{-i\phi_0^{(2)}}|0a\rangle|1\rangle \,,
\end{subeqnarray}
for the first one,
\begin{subeqnarray}
|\Psi_3\rangle &=&|a\Phi_{nc}\rangle|0\rangle,\\ \label{Psi4} 
|\Psi_4\rangle &=&\sin\varphi|a\Phi_{c}\rangle|0\rangle+\cos\psi\cos\varphi \,e^{i(\phi_1^{(1)}-\phi_0^{(2)})}|1a\rangle|0\rangle\nonumber\\
&-&\sin\psi\cos\varphi\,e^{-i\phi_0^{(2)}}|aa\rangle|1\rangle \,,
\end{subeqnarray}
for the second one, and 
\begin{subeqnarray}
|\Psi_5\> &\!\!\!=&\cos\psi\,e^{i\phi_1^{(1)}}|1\Phi_{c2}\>|0\>-\sin\psi|a\Phi_{c2}\>|1\> ,\\
|\Psi_6\> &\!\!\!=&\!\!\!\! \left[ \sqrt{2}\cos\eta\left( \cos\psi\,e^{i\phi_1^{(1)}}|1\Phi_{c3}\>|0\>-\sin\psi|a\Phi_{c3}\>|1\>\right)\right. \nonumber\\
&\!\!\!-&\!\!\!\left.\sin\eta\,e^{-i\phi_0^{(2)}}\!\!\left(\sqrt{2}\cos\psi\,e^{i\phi_1^{(1)}}|1a\>|1\>+\sin\psi|aa\>|2\>\right)\right] \nonumber\\
&\!\!\!/&(1+\cos^2\psi+\sin^2\psi\cos^2\eta),
\end{subeqnarray}
for the third one.

The mixing angles are determined by the Rabi frequencies through the relations
\begin{subeqnarray}
 \tan\eta&=&\Omega^{(2)}/g^{(2)}\\
\tan\psi&=&\Omega_1^{(1)}/g^{(1)}, \ \tan\varphi=\sin\psi/\tan\eta.
\end{subeqnarray}

The dark states $|\Psi_{1,2}\>,|\Psi_{3,4}\>,|\Psi_{5,6}\>$ drive respectively the population of the states $|00\>|0\>,|01\>|0\>$; $|a0\>|0\>,|a1\>|0\>$ and $|10\>|0\>,|11\>|0\>$ in the adiabatic limit.

Since the coupling between the dark states of a same subspace are
respectively of the form
\begin{subeqnarray}
\langle\Psi_2|\frac{d}{dt}|\Psi_1\rangle & = & -\dot{\theta}\cos\eta ,\\
\langle\Psi_4|\frac{d}{dt}|\Psi_3\rangle & = & -\dot{\theta}\sin\varphi ,\\
\langle\Psi_6|\frac{d}{dt}|\Psi_5\rangle & = & \dot{\theta}\sqrt{2}\cos\eta ,
\end{subeqnarray}
the six dark states can evolve freely and independently in the
adiabatic limit under the condition
$\theta\equiv\theta_{0}=const$. This condition can be satisfied when the
amplitudes of the lasers interacting with the second atom vary with a
constant ratio. It guarantees that there is no geometric phase, which would be detrimental for robustness.

The dark states $|\Psi_{1,3}\>$ are stationnary states and do not participate in the dynamics. 
Furthermore, since $g^{(k)}$ is time independent, we remark that for an arbitrary pulse sequence involving $\Omega^{(2)}$ and $\Omega_1^{(1)}$, the initial population of the states $|00\>|0\>,|01\>|0\>,|10\>|0\>,|11\>|0\>$ stays always unchanged at the end of such a process.

\begin{figure}[ht]
\includegraphics[scale=1]{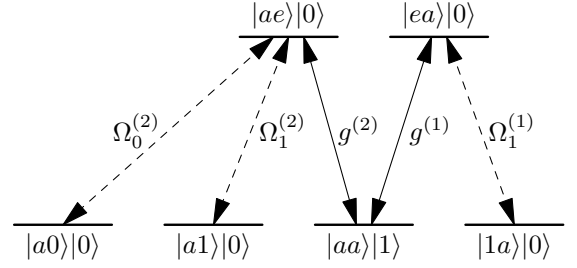}
\caption{Linkage pattern associated to the dark state $|\Psi_{4}\>$. }\label{couplage_sombre}
\end{figure}

The dark state $|\Psi_{4}\>$ is the principal eigenstate involved in the gate operation. It evolves according to the linkage pattern represented in Fig.~\ref{couplage_sombre}.
There are similarities between this linkage pattern and the one of the tripod-type system used for single qubit rotations~\cite{kis}.
It shows that the states $|a0\>|0\>,|a1\>|0\>,|1a\>|0\>$ can respectively evolve in the same way as the states $|0\>,|1\>,|a\>$ involved in the single qubit rotation, leading to a robust rotation of the two-atom states $\left\lbrace |a0\rangle|0\rangle,\,|a1\rangle|0\rangle\right\rbrace$.

We first describe precisely the dynamics of this two-atom states rotation, before applying it to the construction of the arbitrary state controlled-unitary gate. 

\subsection{Robust rotation of two-atom states}
We start with the initial state 
\begin{eqnarray}\nonumber
|\Phi_{i}\rangle&=&\alpha|a1\rangle|0\rangle+\beta|a0\rangle|0\rangle\\ \nonumber
&=&|a\>\otimes(\alpha|1\rangle+\beta|0\rangle)\otimes|0\rangle\\
&=&|a\phi_{i}\>|0\>
\end{eqnarray}
where $\alpha,\beta$ are complex numbers such that $|\alpha|^{2}+|\beta|^{2}=1$.

\paragraph*{-- Step~1\,:} we induce the initial connection to
the dark states
 \begin{align}
|\Phi_{i}\rangle =\alpha_{3}|\Psi_{3}\rangle+\alpha_{4}|\Psi_{4}\rangle
\end{align}
with the constant coefficients
\begin{align}
\begin{cases}
\alpha_{3}=\<\Phi_{nc}|\phi_{i}\>\\
\alpha_{4}=\<\Phi_{c}|\phi_{i}\>
\end{cases}
\end{align}
using the partially overlapping pulse sequence
$\Omega_{1}^{(1)},\Omega^{(2)}$ such that $\varphi$\ decreases from $\pi/2$ to 0. In the adiabatic limit the dark
states evolve independently, such that at the end of the pulse
sequence the statevector becomes
\begin{equation}
|\Phi\rangle=\alpha_{3}|\Psi_{3}\rangle+\alpha_{4}\,e^{i(\phi_{1}^{(1)}-\phi_{0}^{(2)})}|1a\rangle|0\rangle
\end{equation}
since $|\Psi_{3}\rangle$ is a stationary state.

\paragraph*{-- Step~2\,:} we use a pulse sequence in the
reversed order, i.e.
$\Omega^{(2)},\Omega_{1}^{(1)}$ such that $\varphi$ increases from 0 to $\pi/2$. The phases
$\phi_{0,1}^{(2)}$ are unchanged, while we shift by $\delta$ the
phase $\phi_{1}^{(1)}$ of the laser pulse addressing the first qubit. This induces the connection $|1a\>|0\>\rightarrow e^{-i\delta}|\Psi_{4}\>$, and therefore the statevector becomes
\begin{equation}
|\Phi\rangle=\alpha_{3}|\Psi_{3}\rangle+\alpha_{4}\,e^{-i\delta}|\Psi_{4}\rangle,
\end{equation}
and at the end of the pulse sequence
\begin{align}\nonumber
|\Phi_{f}\rangle&=|a\>(\alpha_{3}|\Phi_{nc}\>+\alpha4\,e^{-i\delta}|\Phi_{c}\>)|0\>\\
&=e^{-i\delta/2}|a\>U(\delta,\mathbf{n})|\phi_{i}\rangle|0\>,\label{rotation}
\end{align}
where
\begin{equation}
U(\delta,\mathbf{n})=\exp(-i\dfrac{\delta}{2}\mathbf{n}\cdot\hat{\bm{\sigma}})\label{rotation2}
\end{equation}
is a general rotation of $\hbox{SU}(2)$ of angle $\delta$ around the vector $\mathbf{n}$.
The components of this vector $\mathbf{n}$ are
\begin{equation}
 \mathbf{n}=(\sin2\theta\cos\phi^{(2)},\sin2\theta\sin\phi^{(2)},\cos2\theta),
\end{equation}
and $\hat{\bm{\sigma}}=(\sigma_{x},\sigma_{y},\sigma_{z})$ are the
Pauli operators defined for the second qubit\,:
$\sigma_{x}=|0 \rangle\langle 1|+|1\rangle\langle
0|$, $\sigma_{y}=i(|0\rangle\langle
1|-|1\rangle\langle 0|)$,
$\sigma_{z}=|0\rangle\langle 0|-|1\rangle\langle
1|$.
We notice that the initial population of the
states $|01\rangle|0\rangle$ and $|00\rangle|0\rangle$ ($|10\>|0\>$ and $|11\>|0\>$), which are
connected to the dark states $|\Psi_{1}\rangle$ and
$|\Psi_{2}\rangle$ ($|\Psi_{5}\rangle$ and
$|\Psi_{6}\rangle$), stay unchanged at the end of the process.

The first qubit controls the rotation applied on the second qubit. 
Indeed, equations (\ref{rotation})
and (\ref{rotation2}) show that the process described induces, up to a global phase $-\delta/2$, a rotation of the second qubit of angle $\delta$
around the vector $\mathbf{n}$ on the Bloch sphere
only if the first one is in the ancillary state $|a\>$.
Then, by transferring the
population of an arbitrary preselected state of the first
qubit on state $|a\>$ before realising the two-atom rotation,
we get a controlled-unitary gate generalised to the case of
an arbitrary state for the control qubit.
Moreover, if the couplings between the cavity mode and the atoms are much stronger than the classical laser field interaction, then the cavity is
negligibly populated and the coupling between the atoms are given by a virtual photon. The proposed process is then decoherence-free in the sense that spontaneous radiation from the excited state and cavity damping are avoided.

\subsection{The arbitrary state controlled-unitary gate}
The extension of the previous process to the implementation of the arbitrary state controlled-unitary gate is now simple\,: it consists to transfer as a preliminary step the controlled state $|\phi_{c}\>$ of the first qubit to its ancillary state $|a\>$. 
This can be done using two lasers of appropriate polarisations of Rabi frequencies\,: $\Omega_{a(sti)}^{(1)}$ and
\begin{subeqnarray}
\Omega_{0(sti)}^{(1)}(t)&=&\Omega^{(1)}_{(sti)}(t)\cos\chi\ ,\\
\Omega_{1(sti)}^{(1)}(t)&=&\Omega^{(1)}_{(sti)}(t)\sin\chi\ .
\end{subeqnarray}
The latter of elliptical polarisation is referred to as $\Omega^{(1)}_{(sti)}$.
They drive respectively in a non-resonant way the transition of states $|a\>,|0\>,|1\>$ of the first qubit to $|e\>$ with a one-photon detuning. Alternatively, we prefer to use more efficient one-photon resonant transitions to a second excited atomic state. The important point is to discard the transition $|a\>-|e\>$ by the cavity in this preliminary step.

They define the control-state of the control-qubit and its orthogonal state
\begin{subeqnarray}
|\phi_{c}\>&=&\sin\chi\,e^{i\phi^{(1)}}|1\>+\cos\chi|0\>,\\
|\phi_{nc}\>&=&\cos\chi\,e^{i\phi^{(1)}}|1\>-\sin\chi|0\>.
\end{subeqnarray}
The fuul process is decomposed in three steps\,:
\paragraph*{-- Step~1\,:} we start from a general initial state written in the basis $\left\lbrace |\phi_{nc}\>,|\phi_{c}\>\right\rbrace \otimes\left\lbrace |0\>,|1\>\right\rbrace $
\begin{equation}
|\psi_i\>=|\phi_{nc}\>(\alpha_{1}|0\rangle+\alpha_{2}|1\rangle)|0\rangle
+|\phi_{c}\>(\alpha_{3}|0\rangle+\alpha_{4}|1\rangle)|0\rangle,
\end{equation}
where $\alpha_{i=1,..,4}$ are complex numbers such that $\sum_{i=1}^4|\alpha_i|^2=1$.

We use a f-STIRAP-process with the pulse sequence
$\Omega_{a(sti)}^{(1)},\Omega^{(1)}$ with relative phase
$\xi$ in order to transfer for the first qubit the population of
state $|\phi_{c}\>$ to state $|a\rangle$. This gives the state
\begin{equation}
\label{psi_001}
|\psi_1\rangle=|\phi_{nc}\>(\alpha_{1}|0\rangle+\alpha_{2}|1\rangle)|0\rangle
-e^{i\xi}|a\>(\alpha_{3}|0\rangle+\alpha_{4}|1\rangle)|0\rangle.
\end{equation}
\paragraph*{-- Step~2\,:} we apply the process previously described to implement the rotation $U(\delta,\mathbf{n})$ [see Eqs.~(\ref{rotation},\ref{rotation2})]. The state (\ref{psi_001}) becomes
\begin{align}
|\psi_2\rangle&=|\phi_{nc}\>(\alpha_{1}|0\rangle+\alpha_{2}|1\rangle)|0\rangle\\ \nonumber
&-e^{i\xi}e^{-i\frac{\delta}{2}}|a\>U(\delta,\mathbf{n})(\alpha_{3}|0\rangle+\alpha_{4}|1\rangle)|0\rangle).
\end{align}
\paragraph*{-- Step~3\,:} we make the inverse operation of step~1, i.e. a f-STIRAP-process with the pulse sequence $\Omega^{(1)},\Omega_{a(sti)}^{(1)}$ with relative phase $\xi^{\prime}$ in order to transfer in the first qubit the population of state $|a\rangle$ to state $|\phi_{c}\>$.
The final system state reads
\begin{align}\label{cu}
|\psi_{3}\rangle&=|\phi_{nc}\>(\alpha_{1}|0\rangle+\alpha_{2}|1\rangle)|0\rangle\\ \nonumber
&+e^{i(\xi-\xi^{\prime})}e^{-i\frac{\delta}{2}}|\phi_{c}\>U(\delta,\mathbf{n})
(\alpha_{3}|0\rangle+\alpha_{4}|1\rangle)|0\rangle .
\end{align}

Under the condition $\xi-\xi^{\prime}-\delta/2=0$ the undesirable phase factor of the state~(\ref{cu}) vanishes and one obtains directly the arbitrary state controlled-unitary gate which makes the unitary operation $U(\delta,\mathbf{n})$ on the second qubit only if the first one is in state $|\phi_{c}\>$.

\section{Numerical simulation}\label{Numeric}
\begin{figure}[!ht]
\includegraphics[width=8.6cm]{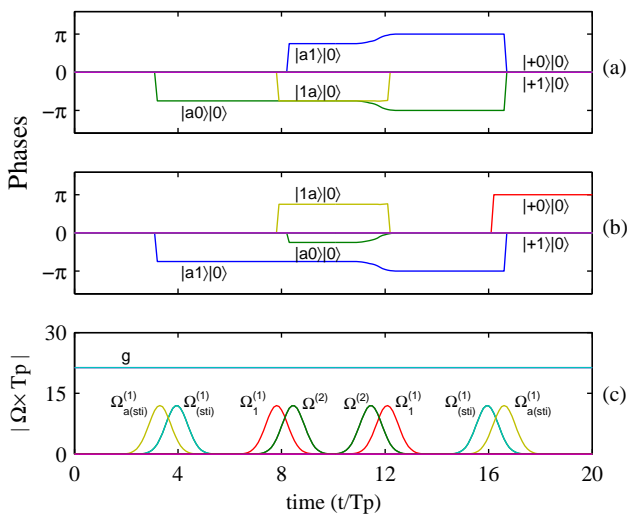}
\caption{(Colour online) Time evolution of the  phases of the probality amplitudes for the initial states $|+0\>|0\>$ (upper frame), $|+1\>|0\>$ (middle frame). The Rabi frequencies are represented in the lower frame. \label{cugphases}}
\end{figure}

\begin{figure}[!ht]
\includegraphics[width=8.6cm]{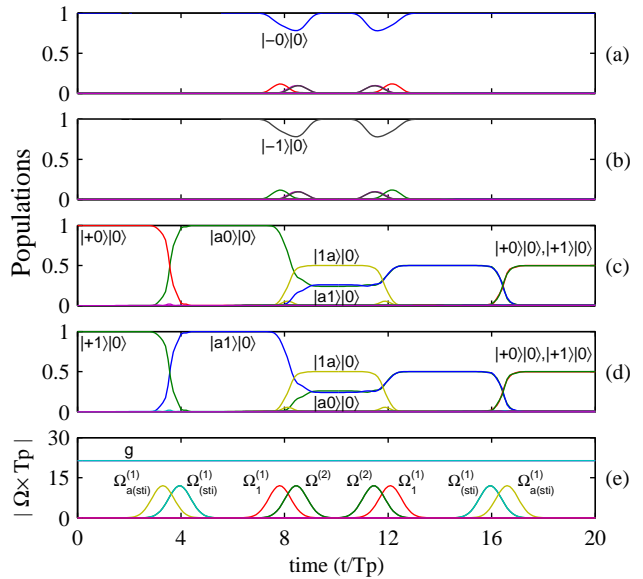}
\caption{(Colour online) Time evolution of the populations represented respectively for the initial states $|-0\>|0\>$, $|-1\>|0\>$, $|+0\>|0\>$, $|+1\>|0\>$ [frames $(a)\hbox{-}(d)$]. The Rabi frequencies are represented in the lower frame.\label{cugpop}}
\end{figure}

We show the numerical simulation of the arbitrary state controlled-unitary gate on Figs.~\ref{cugphases}-\ref{cugpop}.
We have chosen Rabi
frequencies of gaussian shape and of full width at half maximum
$T_P=100\,\hbox{ns}$. The couplings are parametrised by $\Omega_{\max}/2\pi=14\,\hbox{MHz}$ and
$g/2\pi=34\,\hbox{MHz}$ which can be currently obtained experimentally with recent technologies~\cite{Vitanov2,Miller}. The simulation is made for the state of the control qubit $|+\>\equiv\frac{1}{\sqrt{2}}(|0\>+|1\>)$ (its orthogonal state is denoted $|-\>\equiv\frac{1}{\sqrt{2}}(|0\>-|1\>)$).
We have represented in Fig.~\ref{cugphases} the time evolution of the phases associated to the probability amplitudes for the initial states $|+0\>|0\>$, $|+1\>|0\>$. Fig.~\ref{cugpop} exibits the time evolution of the populations for the initial states $|-0\>|0\>$, $|-1\>|0\>$, $|+0\>|0\>$, $|+1\>|0\>$.
They show that when the control qubit is in state $|-\>$, the state of the target qubit is unchanged [Figs.~\ref{cugpop}(a) and~\ref{cugpop}(b)]\,; and when the control qubit is in state $|+\>$, a $R(\pi/4)$ gate is applied on the target qubit, such that its states $|0\>$ and $|1\>$ become respectively $(|0\>+|1\>)/\sqrt{2}$ [Figs~\ref{cugphases}(a) and~\ref{cugpop}(c)] and $(-|0\>+|1\>)/\sqrt{2}$ [Figs~\ref{cugphases}(b) and~\ref{cugpop}(d)].

\section{Discussion and conclusion}\label{End}
The implementaion of the arbitrary state controlled-unitary gate proposed is robust under the adiabatic conditions $\Omega^{(i)}_{j,\,max}T_P,g^{(i)}T_P\gg1$, $\Omega^{(i)}_{j,\,max},g^{(i)}\gg\kappa,1/\tau$ where $\kappa,1/\tau$ are the cavity decay rate and the spectral linewidth of the excited atomic states. It does not involve spontaneous emission since the dynamics follows dark states.
However, since the population of the states $|-0\>|0\>$ and $|-1\>|0\>$ evolves partially along dark states  $|\Psi_{5,6}\>$, which are superpositions of several one-photon states, the coherence of the process is sensitive to the cavity decay rate. These losses are negligible under the condition $g^{(i)}\gg\Omega^{(i)}_{j,\,max}$ where the cavity is negligibly populated, and cavity damping is thus avoided. We have calculated the gate fidelity $\mathcal{F}_-,\mathcal{F}_+$ in Tab.~\ref{Fidelity} for different values of the parameters $(\Omega^{(i)}_{j,max},\ g^{(i)},\ \kappa)$. 
$\mathcal{F}_-,\mathcal{F}_+$ stand respectively for $|\<-_{id}|-_{num}\>|^2$, $|\<+_{id}|+_{num}\>|^2$ where $|\pm_{id}\>,|\pm_{num}\>$ denote respectively the ideal final state without cavity decay and the final state of the numerical simulation for the evolution of an initial state with a control-qubit in state $|\pm\>$, and a target qubit in state $|0\>$ or $|1\>$.

\begin{table}[htb]
\begin{ruledtabular}
\begin{tabular}[c]{cccccr@{,\,}c@{,\,}lccccc}
\multicolumn{8}{c}{$(\Omega^{(i)}_{j,max},\ g^{(i)},\ \kappa)/2\pi$ (MHz)} & $\mathcal{F}_-$ & & $\mathcal{F}_+ $\\ \hline
 & & & & &(14 & 34 & 4.1) & 0.281& & 0.854 & & \\ 
 & & & & &(14 & 34 & 2.05) & 0.488 & & 0.918 & & \\ 
 & & & & &(14 & 34 & 1) & 0.680 & & 0.954 & & \\ 
 & & & & &(14 & 68 & 4.1) & 0.668 & & 0.954 & & \\ 
 & & & & &(14 & 68 & 2.05) & 0.799 & & 0.966  & & \\ 
 & & & & &(14 & 68 & 1) & 0.892 & & 0.976  & & \\
\end{tabular} 
\end{ruledtabular}
\caption{Fidelity of the arbitrary state controlled-unitary gate.\label{Fidelity}}
\end{table} 

We have presented a scheme adapted for the implementation of a universal two qubit quantum gate that generalise the controlled-unitary gate to an arbitrary control state of the first qubit. 
This arbitrary state controlled-unitary gate opens up novel applications for the rapid implementation
of quantum algorithms.
\begin{figure}[!h]
\includegraphics{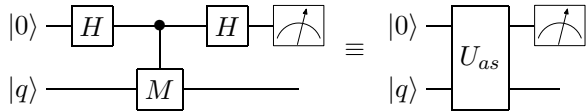}
\caption{Quantum circuit realising a projective measurement and its equivalent using the arbitrary state controlled-unitary gate denoted $U_{as}$. $H$ stands for the Hadamard gate.}\label{circ}
\end{figure}
For instance, the main part of the projective measurement circuit~\cite{Chuang} can be built directly from this gate, as represented in Fig.~\ref{circ}, where $M$ is an unitary operator of eigenvalues $\pm1$. The output qubit of this circuit is an eigenvector of $M$ depending of the result of the measurement of the first qubit. This circuit offers many applications in quantum error corrections~\cite{Chuang,Vala}.

\begin{acknowledgements}
N.S. acknowledges financial supports from the EU network QUACS under contract
No. HPRN-CT-2002-0039 and from La Fondation Carnot.
\end{acknowledgements}


\begin{references}
\bibitem{Bremner} M.~J.~Bremner, C.~M.~Dawson, J.~L.~Dodd, A.~Gilchrist, A.~W.~Harrow, D.~Mortimer, M.~A.~Nielsen, T.~J.~Osborne, Phys. Rev. Lett. {\bf{89}}, 247902 (2002).
\bibitem{Chuang} I.~L.~Chuang and M.~A.~Nielsen, \emph{Quantum Computation and Quantum Information}, Cambridge University Press, Cambridge (2000).
\bibitem{Preskill} J.~Preskill, Proc. R. Soc. London A {\bf{454}}, 385 (1998).
\bibitem{Duan} L.~M.~Duan, I.~J.~Cirac, P.~Zoller, Science {\bf{292}}, 1695 (2001).
\bibitem{bergmann2} K.~Bergmann, H.~Theuer and B.~W.~Shore, Rev. Mod. Phys. {\bf{70}}, 1003 (1998).
\bibitem{vitanov} N.~V.~Vitanov, K.~A.~Suominen and B.~W.~Shore, J. Phys. B {\bf{32}}, 4535 (1999).
\bibitem{unanyan} R.~Unanyan, M.~Fleischhauer, B.~W.~Shore and K.~Bergmann, Opt. Commun. {\bf{155}}, 144 (1998).
\bibitem{kis} Z.~Kis and F.~Renzoni, Phys. Rev. A {\bf{65}}, 032318 (2002).
\bibitem{Goto} H.~Goto and K.~Ichimura, Phys. Rev. A {\bf{70}}, 012305 (2004).
\bibitem{Pellizari} T.~Pellizzari, S.~A.~Gardiner, J.~I.~Cirac and P.~Zoller, Phys. Rev. Lett. {\bf{75}}, 3788 (1995).
\bibitem{Kapale} K.~T.~Kapale, G.~S.~Agarwal, M.~O.~Scully, Phys. Rev. A {\bf{72}}, 052304 (2005).
\bibitem{SingleQG} X.~Lacour, S.~Gu\'erin, N.~V.~Vitanov, L.~P.~Yatsenko, H.~R.~Jauslin, preprint.
\bibitem{Sangouard} N.~Sangouard, X.~Lacour, S.~Gu\'erin, H.~R.~Jauslin, Phys. Rev. A {\bf{72}}, 062309 (2005).
\bibitem{CNOT} N.~Sangouard, X.~Lacour, S.~Gu\'erin, H.~R.~Jauslin, Eur. Phys. J. D DOI: 10.1140/epjd/e2005-00315-2 (2005).
\bibitem{Miller} R.~Miller, T.~E.~Northup, K.~M.~Birnbaum, A.~Boca, A.~D.~Boozer, H.~J.~Kimble, J. Phys. B: At. Mol. Phys. {\bf{38}}, S551 (2005).
\bibitem{Vitanov2} N.~V.~Vitanov, M.~Fleischhaeur, B.~W.~Shore, K.~Bergmann, Adv. Ad. Mol. Opt. Phys. {\bf{46}}, 55 (2001).
\bibitem{Vala} J.~Vala, K.~B.~Walley, D.~S.~Weiss, e-print arXiv:quant-ph/0510021.
%
\end{references}
\end{document}